\documentclass[12pt]{article}
\usepackage{amssymb}
\usepackage{amsmath}
\usepackage{hyperref}
\begin{document}
\title{\bf Conductivity and Shear Viscosity of $arcsin$-Yang-Mills  AdS Black Brane}
\author{Mehdi Sadeghi
	\thanks{Corresponding author: Email:mehdi.sadeghi@abru.ac.ir} 
	 \hspace{2mm} and
      S. M. Moosavi Khansari\thanks{Email:  m.moosavikhansari@abru.ac.ir}\hspace{2mm}\\
{\small {\em Department of Physics, School of Sciences,}}\\
        {\small {\em Ayatollah Boroujerdi University, Boroujerd, Iran}}
       }
\date{\today}
\maketitle
\abstract{ In this paper, a non-abelian $arcsin$-Yang-Mills AdS black brane solution is introduced. Then, the color non-abelian direct current (DC) conductivity and shear viscosity to entropy density ratio of this model is calculated using fluid-gravity duality. Our results show that the Kovtun, Son and Starinets (KSS) bound is saturated and is exactly equal to $\frac{1}{4 \pi}$ but the color conductivity bound is violated for this model. Also, our outcomes recover the Yang-Mills AdS black brane when the coupling of Yang-Mills and gravity fields approaches zero.}\\


\section{Introduction} \label{intro}
 Models of non-linear electrodynamics (NED) have been introduced to address physical problems in various areas of physics. The initial Big Bang singularity and early-time inflation in cosmology, the infinite electromagnetic energy in Maxwell's electrodynamics, and an effective description of the properties of certain condensed matter materials and optical media in solid state are examples of these problems. Some examples of these non-linear electrodynamics were introduced in \cite{Kruglov:2007bh}, \cite{Kruglov:2001dp}, \cite{Kruglov:2007zr}, \cite{Kruglov:2007pg}, \cite{Gaete:2013dta}. In the early stages of the Universe's creation, electromagnetic fields and gravitational fields were very strong, and non-linear effects played a very important role. Therefore, studying non-linear theories is worthwhile. In this paper, we consider the $arcsin$-Yang-Mills as a non-linear model of quantum chromodynamics (QCD) in AdS space-time.\\
The AdS/CFT duality \cite{Maldacena}-\cite{Aharony} states that the weakly coupled gravity theory in an asymptotically anti-de-Sitter space-time is dual to a strongly coupled field theory living on the AdS boundary. The physics of gravitational bulk can be translated into the boundary field theory through the dictionary of this duality.\\
The fluid-gravity duality \cite{Kovtun2012, Bhattacharyya,Rangamani,J.Bhattacharya2014} is resulted from the AdS/CFT duality in the low energy limit. Hydrodynamics is a good description of the theory only in the hydrodynamic limit or low energy limit. Therefore, the theory on the boundary in the AdS/CFT duality is effectively described by hydrodynamics in this limit. The equations of motion up to the first order of derivative expansion for a relativistic conformal charge fluid are stated by conservation laws as follows.    
\begin{align}
	& \nabla _{\mu } J^{\mu } =0\,\,\,\,\,,\,\,\nabla _{\mu } T^{\mu \nu} =0,\\
	& J^{\mu } =n \, u^{\mu }-\sigma T P^{\mu \nu }\partial_{\nu}(\frac{\mu}{T})\nonumber\\
	& T^{\mu \nu } =(\rho +p)u^{\mu } u^{\nu } +pg^{\mu \nu } -\sigma ^{\mu \nu },\nonumber\\
	&\sigma ^{\mu \nu } = {P^{\mu \alpha } P^{\nu \beta } } [\eta(\nabla _{\alpha } u_{\beta } +\nabla _{\beta } u_{\alpha })+ (\zeta-\frac{2}{3}\eta) g_{\alpha \beta } \nabla .u]\nonumber\\& P^{\mu \nu }=g^{\mu \nu}+u^{\mu}u^{\nu}, \nonumber
	\end{align}
 \indent where $n$, $\sigma$, $\eta$, $\zeta $, $\sigma ^{\mu \nu }$ and $P^{\mu \nu }$ are charge density, charge conductivity, shear viscosity, bulk viscosity, shear tensor and projection operator, respectively \cite{Kovtun2012, Bhattacharyya,Rangamani,J.Bhattacharya2014}.\\
Transport coefficients provide us with additional information about the theory on the boundary. In this paper, we calculate the color non-abelian DC conductivity and the ratio of shear viscosity to entropy density by Kubo formula \cite{Son}-\cite{Policastro2001}.\\
There are universal bounds on the shear viscosity to entropy density ratio and conductivity. KSS bound \cite{Policastro2002} ,$\frac{\eta}{s} \geq \frac{1}{4\pi}$, is saturated for Einstein-Hilbert gravity but it is violated for higher derivative gravity \cite{Hartnoll:2016tri},\cite{Sadeghi:2015vaa,Parvizi:2017boc} and massive gravity \cite{Alberte:2016xja},\cite{Baggioli:2020ljz},\cite{Sadeghi:2018ylh}, Horndeski theories\cite{Baggioli:2021ejg} and deformed AdS $5$–Schwarzschild black brane \cite{Ferreira-Martins:2019svk}. The conductivity bound, $\sigma \geq 1$, is violated for breaking of translation symmetry model \cite{Gouteraux:2016wxj}, a specific type of non-linear electrodynamics \cite{Baggioli:2016oqk} and non-abelian Einstein-Born-Infeld AdS black brane\cite{Sadeghi:2021qou}. We want to investigate whether these bounds are preserved in our model or not.\\
By calculating $\frac{\eta}{s}$ and the color non-abelian direct current conductivity in this paper, we aim to study the dual field theory of the non-linear $arcsin$-Yang-Mills AdS black brane.
\section{ Non-linear $arcsin$-Yang-Mills  AdS black brane}
\label{sec2}
\indent The action of Einstein non-linear $arcsin$-Yang-Mills theory is given by ($8\pi G=1$)\cite{Kruglov:2016ezw},
\begin{eqnarray}\label{action}
S=\int d^{4}  x\sqrt{-g} \bigg[R-2\Lambda -\frac{q_1}{q_2}arcsin\big(q_2 \mathcal{F}\big)  \bigg],
\end{eqnarray}
where $R$ is the Ricci scalar, $\Lambda=-\frac{3}{L^2}$ the cosmological constant, $L$ the AdS radius, $\mathcal{F}= F^{(a)}_{\mu \alpha }F^{ (a)\mu \alpha  }$ is Yang-Mills invariant.
$F^{(a)}_{\mu \nu } $ is the field strength of the Yang-Mills potential $A^{(a)}_{\nu }$,
\begin{align} \label{YM}
F^{(a)}_{\mu \nu } =\partial _{\mu } A^{(a)}_{\nu } -\partial _{\nu } A^{(a)}_{\mu } + f^{(a)}_{(b)(c)}[A^{(b)}_{\mu }, A^{(c)}_{\nu }],
\end{align}
in which the structure constants $f^{(a)}_{(b)(c)}$ are equal to 1, $A^{(a)}_{\nu }$'s are the Yang-Mills potentials
where $q_1$ and $q_2$ are generally arbitrary. $q_1$ is dimensional parameter with the dimension of (length)$^4$.\\
For finding the black brane solution of our model, we consider  the following ansatz:
\begin{equation}\label{metric}
ds^{2} =-f(r) dt^{2} +\frac{dr^{2} }{f(r)} +\frac{r^2}{L^2}(dx^2+dy^2).
\end{equation}
The ansatz of gauge potential for our model is cosidered as,
\begin{equation}\label{background}
{\bf{A}} =\frac{1}{\sqrt{2}}h(r)dt\begin{pmatrix}1 & 0 \\ 0 & -1\end{pmatrix}.
\end{equation}
$\sigma_3= \begin{pmatrix}1 & 0 \\ 0 & -1\end{pmatrix}$ is the gauge group of the gauge field which is written in terms of the diagonal generator of the $SL(2,C)$.\\
Einstein's equation is obtained by varying the action (\ref{action}) with respect to the metric $g_{\mu \nu }$ as,
\begin{equation}\label{EOM1}
R_{\mu \nu }-  \tfrac{1}{2} g_{\mu \nu } R + \Lambda g_{\mu \nu }=\kappa q_1 T_{\mu \nu },
\end{equation}
where,
\begin{equation}
 T_{\mu \nu }= \tfrac{1}{2 q_2} arcsin\big(\frac{q_2}{4}\mathcal{F} \big) g_{\mu \nu} - \tfrac{ F_{\mu}^{(a)\alpha } F_{\nu \alpha }{}^{(a)  }}{2\sqrt{1-\frac{1}{16}q_2^2\mathcal{F}^2}}. 
\end{equation}
The non-linear $arcsin$-Yang-Mills' equation is is obtained by varying the action (\ref{action}) with respect to the gauge field $A^{(a)}_{\mu}$,
 \begin{eqnarray}\label{EOM-YM}
\nabla_{\mu}\bigg(\frac{q_1 F^{(a) \mu \nu} }{2 \sqrt{1-\frac{1}{4}q_2^2\mathcal{F}^2}}  \bigg)=0.
\end{eqnarray}
The $(t,t)$ component of Einstein's field equations is determined by utilizing Eq.(\ref{EOM1}) as follows,
\begin{eqnarray}\label{tt}
 f + \Lambda  r^2 + r f' + \frac{q_1 r^2}{2 q_2} arcsin(\frac{1}{2} q_2  h'^2 )-\frac{q_1 h'^2}{\sqrt{4+q_2^2 h'^4}}=0,
\end{eqnarray}
for $q_1=0$ we have,
\begin{equation}\label{ttq2}
 f + \Lambda  r^2 + r f' =0.
\end{equation}
The other component of Einstein's equations is the same as Eq.(\ref{tt}) because there is an unknown function ,$f(r)$, in the ansatz Eq.(\ref{metric}).\\
$h(r)$ is found by using  Eq.(\ref{EOM-YM}) and Eq.(\ref{metric}) as follows,
\begin{eqnarray}\label{h}
h(r)&&= \frac{\sqrt{2}}{Q q_2} \int
^{r}du\sqrt{\sqrt{u^8+Q^4 q_2^2}-u^4}\nonumber\\&&=\frac{\sqrt{2}}{3 Q q_2}\bigg[-2^{\frac{7}{4} }Q\sqrt{q_2} \bigg(\big(r^8+Q^4 q_2^2\big)^{\frac{1}{2}}-r^4\bigg)^{\frac{1}{4}} \nonumber\\&& F\bigg(\frac{1}{8},\frac{3}{4};\frac{9}{8};\frac{(r^4-\sqrt{r^8+Q^4 q_2^2})^2}{Q^4 q_2^2}\bigg)+r\sqrt{\frac{r^8+Q^4 q_2^2}{\sqrt{r^8+Q^4 q_2^2}-r^4}}\nonumber\\&&-\frac{r^5}{\sqrt{r^8+Q^4 q_2^2}-r^4}\bigg]+C,
\end{eqnarray}
and $F(a, b; c; z)$ is the hyper-geometric function.\\
$f(r)$ is found by substituting Eq.(\ref{h}) in Eq.(\ref{tt}) and solving Eq.(\ref{tt}) as below, 
\begin{equation} \label{f}
	f(r)=\frac{2 m}{r}+\frac{ r^2}{L^2}+\frac{q_1}{2r}\int^r u^2\bigg(\frac{2  h'(u)^2}{\sqrt{4+q_2^2 h'(u)^4}}-\frac{1}{q_2} arcsin\big(\frac{1}{2}q_2 h'(u)^2\big)\bigg)du,
\end{equation}
where,
\begin{equation} 
	h'(u)=\frac{\sqrt{2}}{Q q_2}\sqrt{\sqrt{u^8+Q^4 q_2^2}-u^4}.
\end{equation}
\section{Conductivity and Shear Viscosity}
\label{sec3}
\indent The AdS/CFT duality can be used to describe strongly coupled non-perturbative quantum systems in terms of weakly coupled gravity quantities. For example, a field $\phi$ in gravity is dual to an operator $\mathcal{O}_{\phi}$ on the CFT side. The conductivity is related to the retarded correlation function of currents at zero momentum through the Green-Kubo formula \cite{Son} as follows,
\begin{equation} \label{kubo2}
	\sigma^{ij}_{ab} (k_{\mu})=\frac{1}{\omega}\int dt \theta(t) e^{i \omega t}<[J^i_a(t, \bold{K}),J^j_b(t, \bold{K})]>=-\mathop{\lim }\limits_{\omega \to 0} \frac{1}{\omega } \Im G^{ij}(k_{\mu}).
\end{equation}
So the retarded two-point current correlator of CFTs at finite temperature is required for calculating conductivity.\\
The gauge potential, $A^{(a)}_{\mu}$, on the bulk theory is dual to the current, $J^i_{\mu}$, on the boundary theory. Therefore, the gauge potential must be perturbed as $\tilde{A}_x=\tilde{A}_x(r)e^{-i\omega t}$ on bulk of AdS. Where $a,b$ indices refer to $SL(2,C)$ group symmetry.\\
We insert the gauge potential perturbation into the action Eq.(\ref{action}) and expand the resulting action up to the second order of the perturbed part. We have,
 \begin{align}\label{action-2}
 S^{(2)}=-\int d^4x \frac{4 q_1 }{ f \sqrt{1-q_2^2 h'(r)^4}} \Bigg[-f^2 \left((\partial_r\tilde{A}_x^{(1)})^2+(\partial_r\tilde{A}_x^{(2)})^2+(\partial_r\tilde{A}_x^{(3)})^2\right) \nonumber\\+
  \big(\omega^2+h^2\big) \bigg((\tilde{A}_x^{(1)})^2+(\tilde{A}_x^{(2)})^2\bigg)+\omega^2(\tilde{A}_x^{(3)})^2  \Bigg].
 \end{align}
The equations of motion for perturbed gauge fields are obtained by varying the action $S^{(2)}$ (\ref{action-2}) with respect to $\tilde{A}_x ^{(1)}$, $\tilde{A}_x ^{(2)}$, $\tilde{A}_x ^{(3)}$, respectively, as the following,\\
 \begin{align}\label{PerA1}
 f \left(f\tilde{A}_x^{(1)'}\right)'+ \tilde{A}_x^{(1)}\left( h^2+ \omega ^2\right)+\frac{2 q_2^2 f(r)^2 h'(r)^3 h''(r) \tilde{A}_x^{(1)'}}{1-q_2^2 h'(r)^4} =0.     
 \end{align}
 \begin{align}\label{PerA2}
	f \left(f\tilde{A}_x^{(2)'}\right)'+ \tilde{A}_x^{(2)}\left( h^2+ \omega ^2\right)+\frac{2 q_2^2 f(r)^2 h'(r)^3 h''(r) \tilde{A}_x^{(2)'}}{1-q_2^2 h'(r)^4} =0.     
\end{align}
\begin{align}\label{PerA3}
	f \left(f\tilde{A}_x^{(3)'}\right)'+ \omega ^2\ \tilde{A}_x^{(3)}-\frac{2 q_2^2 f(r)^2 h'(r)^3 h''(r) \tilde{A}_x^{(3)'}}{1-q_2^2 h'(r)^4} =0.     
\end{align}
Eq.(\ref{PerA1}) is the same as Eq.(\ref{PerA2}).\\
Now we need to find the solutions of Eq.(\ref{PerA1}) and Eq.(\ref{PerA3}). First, we solve them near the event horizon.\\
Since the condition $\tilde{A}_x^{(a)}(r_h)=0$ holds, and given the relation $f\sim f'(r_h)(r-r_h)$, we consider the following ansatz:\\
 \begin{align}
 \tilde{A}_x^{(a)}\sim (r-r_h)^{z_a} \, , \qquad a=1,2,3
 \end{align}
 where,
 \begin{eqnarray}\label{z12}
 &z_1=z_2=\pm i \frac{ \sqrt{\omega^2+h(r_h)^2} }{4 \pi T},\\
 &\label{z3}
 z_3=\pm i \frac{\omega   }{4 \pi T }.
 \end{eqnarray}
The Hawking temperature of the black brane $T$ is given as follows,
\begin{equation}
T=\frac{1}{2 \pi} \Big[ \frac{1}{\sqrt{g_{rr}}}\frac{d}{dr}\sqrt{-g_{tt}}\Big]\Bigg|_{r=r_h}=\frac{ f'(r_h)}{4 \pi}.
\end{equation} 
 For general solution of $\tilde{A}_x^{(a)}$, we consider the following ansatz, 
 \begin{align}\label{EOMA1}
 \tilde{A}_x^{(1)}=\tilde{A}^{(1)}_{\infty}\Big(\frac{-3f}{\Lambda r^2}\Big)^{z_1}\Big(1+i\omega b_1(r)+\cdots\Big) ,
 \end{align}
 \begin{align}\label{EOMA2}
 \tilde{A}_x^{(2)}=\tilde{A}^{(2)}_{\infty}\Big(\frac{-3f}{\Lambda r^2}\Big)^{z_2}\Big(1+i\omega b_2(r)+\cdots\Big) ,
 \end{align}
 \begin{align}\label{EOMA3}
 \tilde{A}_x^{(3)}=\tilde{A}^{(3)}_{\infty}\Big(\frac{-3f}{\Lambda r^2}\Big)^{z_3}\Big(1+i\omega b_3(r)+\cdots\Big) ,
 \end{align}
 where $\tilde{A}^{(a)}_{\infty}$ represents the value of fields at the boundary, $z_i$'s correspond to the negative signs in Eq.(\ref{z12}) and Eq.(\ref{z3}) and $r$ ranges from the event horizon to the boundary.\\
  $b_3(r)$  in Eq.(\ref{EOMA3}) is as follows,
\begin{equation}\label{b3}
b_3(r)=C_3+\int^r\frac{\left(C_4+\int^{u_2}Y(u_1)du_1\right) \sqrt{1-q_2^2 h'(u_2)^4}}{f(u_2)}du_2,
\end{equation}
 where $C_3$ and $C_4$ are integration constants and $Y(u_1)$ is as follows,
 \begin{equation} 
Y(u_1)= -\frac{u_1^2 f''(u_1)-2 u_1 f'(u_1)+2 f(u_1) }{u_1^2 \left(1-q_2^2
 	h'(u_1)^4\right)^{1/2}}+\frac{2
 	q_2^2 u_1 \left(u_1 f'(u_1)-2 f(u_1)\right) h'(u_1)^3 h''(u_1)}{u_1^2 \left(1-q_2^2
 	h'(u_1)^4\right)^{3/2}}.
\end{equation}
 The solution should be regular on the event horizon. Therefore, $C_{4}$ is determined by demanding the application of regularity to $h_3(r)$ at the horizon as,
  \begin{equation}
 C_{4}=-\int^{r_h}Y(u_1)du_1.
 \end{equation} 
 Considering the solution of $\tilde{A}_x^{(3)}$ in Eq.(\ref{action-2}) and the variation with respect to $\tilde{A}^{(3)}_{\infty}$, Green's function \cite{Son} can be obtained using Witten's prescription \cite{Witten:1998qj} as follows,
\begin{align} \label{Green1}
 & G_{xx}^{(33)} (\omega ,\vec{0})=-i \omega \frac{r f'(r)+f(r) \left(r b_3'(r)-2\right)}{4 \pi  r T}\bigg|_{r \to \infty}.
\end{align}
 The conductivity is as following,
 \begin{eqnarray}
 \sigma_{xx}^{(33)}=\frac{r f'(r)+f(r) \left(r b_3'(r)-2\right)}{4 \pi  r T},
  \end{eqnarray}
by substituting $b_3(r)$ in above equation and using Eq.(\ref{kubo2}) we have, 
\begin{eqnarray}
 &&\sigma_{xx}^{(33)} =\sqrt{1-q_2^2 h'(r_h)^4}.
\end{eqnarray}
 It shows that the conductivity bound \cite{Donos:2014cya}  is violated for the non-linear $arcsin$-Yang-Mills  AdS black brane theory. In the limit of $q_2 \to 0$,  we have,
\begin{eqnarray}
\sigma_{xx}^{(33)} =1.
\end{eqnarray}
It can be easily shown that $\sigma_{xx}^{(11)}=\sigma_{xx}^{(22)} =0$.\\
The ratio of shear viscosity to entropy density is another important quantity in fluid-gravity duality. This ratio is proportional to the inverse squared coupling of the theory on the boundary. We want to calculate this ratio using the formula introduced in the \cite{Hartnoll:2016tri}.\\
\begin{equation}
\frac{\eta}{s}=\frac{1}{4 \pi} \phi(r_h)^2,
\end{equation}
where $\phi$ represents the perturbed part of the metric.\\
Therefore, we perturb the metric as $g_{\mu \nu} \to g_{\mu \nu} +\frac{r^2}{L^2} \phi(r )$ \footnote{The components of $t$ and $z$ in $\phi(t,r,z)$ are disappeared by Fourier transforms.} and put it into action, expanding the action up to the second order of $\phi$. Finally, the equation of motion of $\phi(r )$ is obtained by varying the resulting action with respect to $\phi(r)$ as,
\begin{eqnarray}
\phi ' f'+f \phi ''=0.
\end{eqnarray}
The solution of $\phi(r)$ is as follows,
\begin{align}\label{phi}
\phi (r)=C_5 + C_6\int^{r}\frac{1 }{ f(u)}du.
\end{align}
We consider solution of $\phi(r)$ near the event horizon as the following,
\begin{align}
	\phi (r)=C_5+\frac{C_6}{4 \pi T} \log(r-r_h).
\end{align}
By demanding the regularity of $\phi (r)$ on the event horizon. We have, 
\begin{eqnarray}
C_6=0.
\end{eqnarray}
By applying normalization on solution Eq.(\ref{phi}) so $C_5=1$, we have,
\begin{align}
\phi(r)=1.
\end{align}
Finally, the shear viscosity to entropy density ratio is given by,
\begin{equation}
\frac{\eta}{s}=\frac{1}{4 \pi} \phi(r_h)^2=\frac{1}{4 \pi}.
\end{equation}
The $\frac{\eta}{s}$ is satisfied for dual to strongly correlated systems in this universal relation as $\frac{\eta}{s} \geq \frac{1}{4 \pi}$.\\
Our results show that the KSS bound is saturated for the non-linear $arcsin$-Yang-Mills AdS black brane.

\section{Results and discussion}
We introduced a non-abelian $arcsin$-Yang-Mills black brane in AdS spacetime. Conductivity and $\frac{\eta}{s}$ are two important quantities that are calculated via AdS/CFT duality. Since the shear viscosity to entropy density ratio is proportional to the inverse square of the coupling. We conclude that the coupling of this duality in our model is the same as the field theory dual of Einstein-Hilbert gravity, and the conductivity bound, $\sigma \geq 1$ {\footnote{ We consider $\frac{1}{e^2}=1$}, is violated in our model. It indicates the existence of non-linear effects on this bound. Regarding conductivity, it is important to emphasize that the probe limit is assumed. Otherwise, in this model, the DC conductivity would be infinite, and no bounds would be violated at all. We mention the bounds in real fluids \cite{Trachenko:2020jgr} as an application example to convince ourselves that this is not a holographic game. Also, these quantities can be confirmed as experimental data for string theory.

 \section{Conclusions}

\noindent In this paper, the $\frac{\eta}{s}$ and $\sigma$ of the non-linear $arcsin$-Yang-Mills AdS black brane are calculated using fluid-gravity duality. We conclude some information about our field theory dual of this solution. The shear viscosity to entropy density ratio tells us that the coupling of the field theory side is the same as the dual of Einstien-Hilbert gravity on AdS spacetime. Also, the conductivity bound is violated for current model. Our results show that the non-abelian Ohm's law is diagonal and the conductivity is scalar from spatial and gauge group indices.In the following, we mention some future works on this model.\\
This model can be studied in different theories of gravity such as Horendeski, massive gravity, scalar-tensor gravity, bigravity and higher derivative gravities. Hydrodynamic aspects and stability of these solutions can be discussed.\\
The solution of this model in higher dimensions is an interesting subject and the shadow of this solution can be investigated.\\
Considering the bounds on thermal conductivity \cite{Grozdanov:2015djs} and thermal diffusivity \cite{Blake:2016sud},\cite{Baggioli:2017ojd} for this model are another interesting subjects, too.

\vspace{1cm}
\noindent {\large {\bf Acknowledgment} } Authors would like to thank Reza Saadati, Komeil
Babaei, Mojtaba Shahbazi, Farmarz Rahamni and the referee of INJP for useful comments and suggestions.


\end{document}